\documentclass[twocolumn,english,aps,pre]{revtex4-1}
\usepackage[T1]{fontenc}
\usepackage[latin9]{inputenc}
\usepackage{array}
\usepackage{multirow}
\usepackage{amsmath}
\usepackage{graphicx}

\makeatletter

\providecommand{\tabularnewline}{\\}

\makeatother

\usepackage{babel}
\begin{document}

\title{Jamming on curved surfaces}

\author{Christopher J. Burke}

\affiliation{Department of Physics and Astronomy, Tufts University, 574 Boston
Avenue, Medford, Massachusetts 02155, USA}

\author{Timothy J. Atherton}

\affiliation{Department of Physics and Astronomy, Tufts University, 574 Boston
Avenue, Medford, Massachusetts 02155, USA}
\begin{abstract}
Colloidal and other granular media experience a transition to rigidity
known as jamming if the fill fraction is increased beyond a critical
value. The resulting jammed structures are locally disordered, bear
applied loads inhomogenously, possess the minimal number of contacts
required for stability and elastic properties that scale differently
with volume fraction to crystalline media. Here the jamming transition
is studied on a curved ellipsoidal surface by computer simulation,
where shape evolution leads to a reduction in area, crowding the particles
and preventing further evolution of the surface. The arrested structures
can be unjammed and the surface further evolved iteratively, eventually
leading to a rigid metric-jammed state that is stable with respect
to motion of the particles and some specified space of deformations
of the manifold. The structures obtained are compared with those obtained
in flat space; it is found that jammed states in curved geometries
require fewer contacts per particle due to the nonlinearity of the
surface constraints. In addition, structures composed of soft particles
are compressed above the jamming point. It is observed that relatively
well-ordered but geometrically frustrated monodispersed packings share
many signatures of disordered bidispersed packings.
\end{abstract}
\maketitle

\section{Introduction}

Jamming is a transition to rigidity that occurs at high density, low
temperature and low applied stress in granular media, glasses and
foams\cite{Liu2010}. Jammed structures are mechanically stable and
behave as a solid in the bulk, but lack crystalline order. Particularly
well-defined is the jamming point J that occurs at a critical packing
density $\phi_{c}$ at zero temperature and stress; at this point,
packings are stable with respect to an infinitesimal applied stress. 

Packings at the jamming point possess a number of remarkable features:
first, are typically disordered according to some measure of local
crystalline order\cite{Kansal2002}. In two dimensions, this is quantified
by the hexatic order parameter,
\begin{equation}
\psi_{6}=\left\langle \exp(i6\theta)\right\rangle ,
\end{equation}
where the average is taken over the nearest neighbors. Second, they
are \emph{isostatic}, or possess the minimal number of contacts required
for mechanical stability\cite{Moukarzel1998}. This number in Euclidean
space is calculated by matching the number of degrees of freedom,
$ND$ for a packing of $N$ spheres in $D$ dimensions, to the number
of constraints $ZN/2$ if each particle has, on average, $Z$ contacts
with neighboring particles. This yields $Z=2n$, i.e. $Z=4$ for 2D.
In contrast, a crystalline hexagonal packing has $Z=6$. Third, if
the particles themselves are deformable, the structure can be compressed
beyond the jamming point. These disordered jammed packings exhibit
a number of scaling laws\cite{OHern2002,OHern2003,Drocco2005,OlsonReichardt2010}.
In particular, the scaling of elastic moduli with density is found
to be very different from that of crystalline solids, due to the abundance
of soft modes. 

Powerful theoretical tools have been developed to classify the nature
of the jammed structures. For hard particles, Torquato and Stillinger\emph{\cite{Torquato2001}
}proposed a taxonomy of jamming based on the space of feasible motions
available to the constituent particles: A packing is \emph{locally
jammed}, the least stringent category, if no particles are able to
move while the others remain fixed; it is \emph{collectively jammed}
if no subset of particles is movable with the remainder held in place;
it is \emph{strictly jammed} if no collective subset of the particles
can be moved at the same time as a volume conserving deformation of
the container. The category to which a configuration belongs can be
determined numerically by solving a linear program, for which efficient
algorithms exist permitting the classification of structures with
large numbers of particles\cite{Donev2004b}. 

In contrast to hard particle packings, packings of particles with
soft, finite range potentials can be compressed beyond the jamming
point. For disordered packings (typically bidispersed packings in
2D or 3D, or monodipsersed packings which are able to avoid crystallization
in 3D), as the density is increased above the jamming point, a number
of specific mechanical properties are observed which distinguish them
from crystalline packings. They show an excess of low-frequency vibrational
modes (in contrast to the Debye law behavior of ordered solids\cite{Kittel}),
as seen in a number of glassy and disordered systems\cite{OHern2003,Silbert2005,Silbert2009,Sussman2015}.
They also exhibit critical scaling laws as the density is increased
above the jamming point\cite{OHern2003}. This is true for the contact
number $Z$, as well as the bulk modulus $B$ and the shear modulus
$G$. These scaling properties reveal the nonlinear nature of packings
near the jamming point. In flat space, monodispersed packings in 2D
are highly crystalline \textemdash{} disordered monodispersed packings
can be produced only under extreme circumstances\cite{Atkinson2014}.
However, for packings in a curved 2D space, crystalline order is frustrated
by the surface geometry, necessitating defects and inducing strain
in the crystalline regions of the packing\cite{Seung1988,Bowick2000,Irvine2010,Bruss2012,Burke2015}.
The question arises, then, of how geometric frustration affects the
mechanical properties at the jamming point.

In this work, we consider the situation where jamming occurs on a
curved surface. One situation where this might happen is if colloidal
particles are trapped by surface tension on the surface of an emulsion
droplet and either the size of the droplet is reduced or the shape
deformed to one with lower surface area. At sufficient density, the
particles become crowded and arrest further evolution of the surface.
This mechanism can stabilize a variety of arrested shapes, including
bispheres\cite{Pawar2011} and ellipsoids\cite{Burke2015}. The questions
we address in this paper are: 1) are these states jammed? and 2) how
does curvature affect the mechanical properties near the jamming point?
Packings of particles on a curved surface are also important in other
applications, for example in spherical codes, i.e. packings of particles
on the surface of a sphere of arbitrary dimensions, which are useful
as error correcting codes\cite{Cohn2011}.

The paper is organized as follows: first studying hard particles,
in section \ref{subsec:Unjamming-arrested-packings} we adapt the
linear program of ref. \cite{Donev2004b} to curved spaces; in section
\ref{subsec:Are-arrested-packings} we apply it to arrested packings,
allowing us to evolve them toward an ultimately arrested state and
compare their features with jammed configurations in Euclidean space;
we examine in section \ref{subsec:Criteria-for-isotaticity} whether
the packings are isostatic and find that the criteria for this must
be modified in the presence of curvature. Packings of soft particles
on curved surfaces are studied in section \ref{subsec:Soft-particles};
we compare the properties of disordered bidispersed packings to relatively
well ordered but geometrically frustrated monodispersed packings.
Brief conclusions are presented in section \ref{sec:Conclusion}.

\section{Results}

\subsection{Unjamming arrested packings\label{subsec:Unjamming-arrested-packings}}

We simulate $N$ hard particles of radius $r$ that move diffusively
on a prolate ellipsoidal surface of fixed volume and aspect ratio
$a$ that evolves from $a_{0}>1$ toward the spherical state $a=1$
as the simulation progresses. The centroids of the particles are fixed
rigidly to the surface and, as the surface evolves, particles are
kept on it using constraint forces imposed through a Lagrange multiplier.
Overlaps caused by surface evolution are resolved by gradient descent.
If the initial coverage of particles is sufficiently high, the particles
become crowded and, at some aspect ratio $a_{f}$, arrest further
evolution of the surface. The algorithm features adaptive time-stepping
to resolve the dynamics near the point of arrest and further details
are given in the Methods section. We refer to the output of this algorithm
as an \emph{arrested} structure; the central question of this section
is to determine whether they are also jammed. The arrested structures
studied here all have an aspect ratio near $a=4$.

To test for jamming, we adapt the linear program developed by Donev
\emph{et. al.}\cite{Donev2004b}, which aims to find a prototypical
unjamming motion $\Delta\mathbf{x}$ by applying a random force $\mathbf{F}$
to the packing. The force $\mathbf{F}_{i}$ experienced by particle
$i$ is drawn from a spherically symmetric gaussian distribution of
unit variance. The program maximizes the virtual work,
\begin{equation}
\mathrm{\underset{\Delta\mathbf{x}}{max}\:\mathbf{F}^{\mathrm{T}}\cdot\Delta\mathbf{x}},
\end{equation}
subject to an interpenetrability constraint for each pair of particles
$(i,j)$,
\begin{equation}
(\Delta\mathbf{x}_{i}-\Delta\mathbf{x}_{j})^{\mathrm{T}}\cdot\hat{\mathbf{x}}_{ij}\ge2r
\end{equation}
where $\hat{\mathbf{x}}_{ij}$ is a unit vector pointing from the
center of particle $i$ to particle $j$, and also subject to an overall
bound on the motion,
\begin{equation}
\left\Vert \Delta\mathbf{x}\right\Vert <X_{max}.
\end{equation}
The appropriate surface constraint is obtained from the level set
representation of the surface $f(\mathbf{x})=0$ by MacLaurin expansion,
\[
f(\mathbf{x}+\Delta\mathbf{x})=f(\mathbf{x})+\frac{\partial f}{\partial\mathbf{x}}\cdot\Delta\mathbf{x}+...=0,
\]
and noting that $\frac{\partial f}{\partial\mathbf{x}}$ is parallel
to the local surface normal $\mathbf{N}$. Hence, the linearized surface
constraints are,
\begin{equation}
\mathbf{N}_{i}\cdot\Delta\mathbf{x}_{i}=0
\end{equation}
where $\mathbf{N}_{i}$ is the surface normal at the $i$th particle
center and corresponding to allowing motion of the particle in the
local tangent plane. Finally, because ellipsoids possess a trivial
rotational symmetry, we also constrain the angular momentum of the
motion $\Delta\mathbf{x}$ about the symmetry axis $\mathbf{\hat{z}}$
to suppress the corresponding motion,
\begin{equation}
\sum_{i}\mathbf{\hat{z}}\cdot(\mathbf{x}_{i}\times\Delta\mathbf{x}_{i})=0.
\end{equation}

We note that the use of a linear constraint for the surface is hence
a more egregious approximation than, say, the interpenetrability condition
because, unless the surface happens to be locally flat, any finite
$\Delta\mathbf{x}$ takes the particle away from the surface and hence
violates the true nonlinear constraint. To compensate for this, we
restrict the bound on the motion $X_{max}\le r$, and, having found
an unjamming motion, test that it is feasible within the nonlinear
constraint. 

\begin{figure}
\includegraphics{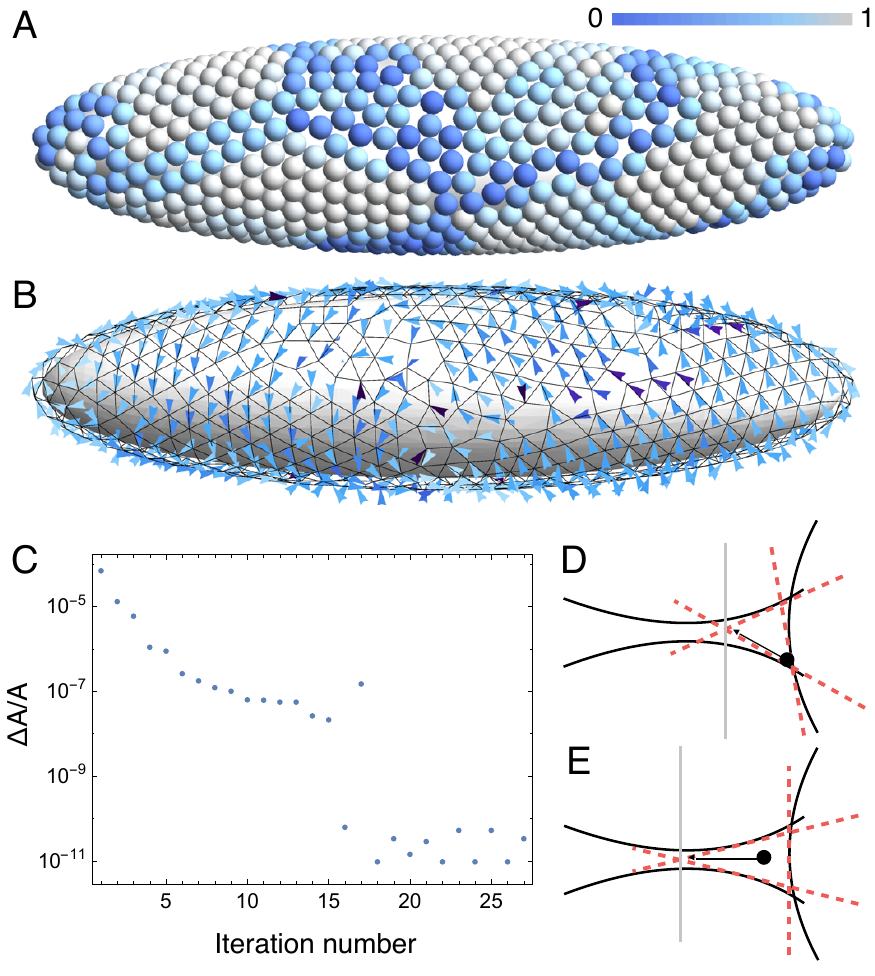}

\caption{\textbf{\label{fig:A-An-arrested}A} An arrested structure produced
by evolution of an ellipsoidal surface at constant volume with the
particles colored by the hexatic order parameter $\psi_{6}$. \textbf{B}
Unjamming motion found for this structure using the linear program.
Executing the unjamming motion permits further evolution of the surface.
\textbf{C} Plot of relative area decrease in successive unjamming
and evolution steps. \textbf{D} Arrested packings often have particles
at the edge of the feasible region of moves, the jamming polytope,
for a particle (solid black line) for which the linear approximant
(red dashed lines) is relatively poor. \textbf{E} Preconditioning
the configuration through gradient descent moves particles to the
center of the jamming polytope, for which the linear approximant has
greater fidelity. }
\end{figure}

A typical arrested configuration is depicted in fig. \ref{fig:A-An-arrested}.
In fig. \ref{fig:A-An-arrested}A, the particles are colored by the
hexatic order parameter $\psi_{6}$, revealing large regions of approximately
crystalline order separated by disordered regions.  Applying the
linear program described above to this configuration reveals the prototypical
unjamming motion depicted in fig. \ref{fig:A-An-arrested}B. Executing
this motion along the surface opens gaps in the packing. By alternating
application of the linear program to unjam the system with further
relaxation of the surface, it is possible to evolve the system further
towards the equilibrium state. Fig. \ref{fig:A-An-arrested}C shows
a typical example of the change in area as a function of iteration
number, showing that after a few applications of the linear program,
the system arrives at an \emph{ultimately arrested }state beyond which
further evolution is precluded. 

We found that the process of successive unjamming and relaxing could
be accelerated by a preconditioning step in which an artificial energy
functional is minimized with respect to the particle positions by
gradient descent. The functional used supplements the hard particle
constraint with a short range repulsive pairwise auxiliary potential,
\begin{equation}
V=\begin{cases}
\infty & x\le2r\\
(x-x_{c})^{2}\left(\log(x-2r)-\log(x_{c})\right) & 2r<x<x_{c}\\
0 & x_{c}\le x
\end{cases}
\end{equation}
where $x$ is the center-to-center particle distance and $x_{c}=2.1r$
is a distance cutoff beyond which particles do not interact. Note
that the logarithmic form of this auxiliary potential diverges as
particles come in contact with one another, preserving the hard-particle
constraint. Preconditioning the particle positions by gradient descent
of the auxiliary potential enables the linear program to more effectively
find unjamming motions. 

The reason for the improvement due to the preconditioning steps may
be understood by considering the set of feasible motions available
to the particles, referred to in the literature as the\emph{ jamming
polytope}\cite{Torquato2010a}. As shown in fig. \ref{fig:A-An-arrested}D,
the output of the arrest algorithm tends to produce packings with
particles that are locally close to the edge of the jamming polytope,
where the linearized version of the jamming polytope is a poor approximation
and artificially constricted. Preconditioning by gradient descent
tends to move particles into the center of the jamming polytope where
the linear approximation is better and hence allows the linear program
to more effectively find an unjamming motion, as in fig. \ref{fig:A-An-arrested}E. 

In addition to the linear program, a full energy minimization of the
auxiliary potential can be performed and often leads to an unjamming
motion. However, this is computationally expensive and thus full energy
minimization steps are only attempted when the linear program fails
to uncover an unjamming motion.

\subsection{Are arrested packings jammed?\label{subsec:Are-arrested-packings}}

\begin{figure}
\begin{centering}
\includegraphics[width=1\columnwidth]{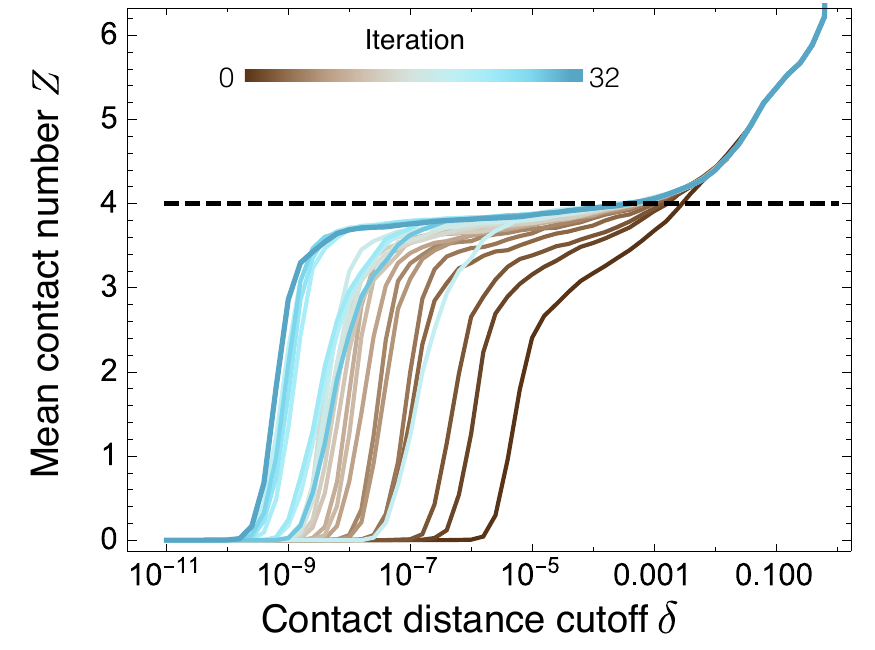}
\par\end{centering}
\caption{\label{fig:Contact-number}Mean number of contacts $Z$ as a function
of the contact cutoff distance $\delta$. Note that the leftward shift
is not monotonic: this is due to the alternating use of the linear
program and the full energy minimization; the energy minimization
results in more even spacing and thus larger interparticle distance
on average.}
\end{figure}

We now wish to determine whether the arrested or ultimately arrested
states resemble a jammed state. A powerful tool to do so is to examine
the contact network. Two particles are defined to be in contact at
a cutoff distance $\delta$ if the distance between them $x<2r+\delta$.
A plot of the average number of contacts $Z$ as a function of the
cutoff distance possesses a characteristic shape for a jammed state\cite{Donev2005}:
$Z=0$ for low $\delta$; at some characteristic \emph{contact lengthscale}
$\delta_{0}$, $Z$ quickly rises to the isostatic value; once $\delta\ge r$
$Z$ diverges like $\delta^{D}$ where $D$ is the dimensionality
of the space. 

We show in fig. \ref{fig:Contact-number} the contact number distribution
$Z(\delta)$ for a sequence of successive preconditioning, unjamming
and surface evolution steps described above. After the initial unjamming
step, the contact lengthscale is $\delta_{0}\sim10^{-5}$. Following
unjamming and evolution, $\delta_{0}$ decreases and approach a limiting
curve with $\delta_{0}\sim10^{-9}$. 

For this ultimately arrested configuration, the linear program is
no longer able to find a finite unjamming motion. The state is therefore
at least \emph{collectively jammed} according to the taxonomy of \cite{Torquato2001},
but this classification is too permissive because it does not account
for deformations of the manifold on which the packing is embedded.
The more restrictive epithet of strictly jammed is not directly applicable
here, because the jamming occurs on a compact surface without boundaries
to deform. Rather, the packing is also jammed with respect to volume
conserving and area-reducing deformations of the ellipsoid; the jamming
is caused by change in the underlying metric of the space as surface
relaxation occurs. 

We therefore propose to call the ultimately arrested configurations
\emph{metric jammed }states, which we define to mean that the packing
is jammed with respect to simultaneous collective motions of the particles
and some well-defined set of deformations of the manifold. This restriction
parallels that of strict jamming, which does not permit arbitrary
deformations of the boundary, but rather requires the allowed deformation
to conserve volume. Hence, the initially arrested states are not jammed,
but rather evolve toward the metric jammed state upon repeated application
of the linear program to unjam them and relaxation of the surface.

\subsection{Criteria for isotaticity\label{subsec:Criteria-for-isotaticity}}

We now compare features of the metric jammed state to those of jammed
states in Euclidean space. Most notably, the contact number distributions
$Z(\delta)$, even in the limit of metric jamming, lack the clear
plateau observed in \cite{Donev2005}. This is because jammed packings
contain some fraction of underconstrained particles that are free
to move within a cage created by their neighbors. These are referred
to as \emph{rattlers} and should be excluded from the average used
to create the plot of $Z(\delta)$\cite{Donev2004b}. 

Rattlers are identified based on the number of contacts they make
with their neighbors \textemdash{} a particle needs three contacts
which are not in the same hemisphere to be constrained. To count contacts
for the purpose of identifying rattlers, we again choose a cutoff
separation distance below which particles are in contact, but we consider
the separation between a particle an its neighbors from every point
available to a particle while its neighbors are held fixed. We do
this because two particles may be close in the simulation output,
but one of the particles may be able to move a significant distance.
For example, a rattler may be close to a neighbor and may be identified
as being in contact with that neighbor, while that contact does not
contribute to the mechanical stability of the packing. We give details
of the contact identification process in the supplementary material.

\begin{figure}
\includegraphics{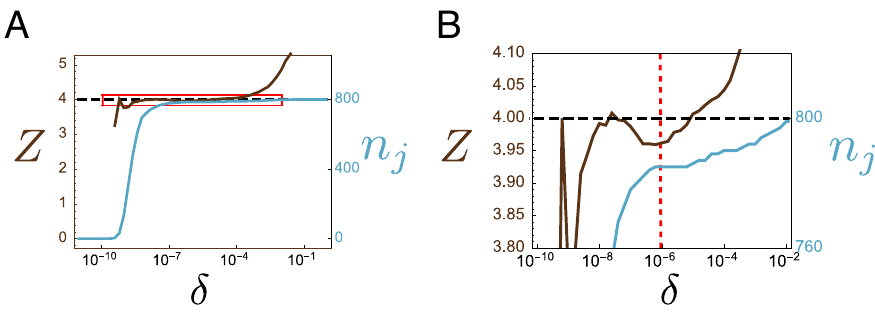}

\caption{\label{fig:contactsWithoutRattlers}A) Plot of Z versus $\delta$
with rattlers removed (brown), alongside the number of non-rattlers
$n_{j}$, i.e. the number of particles in the packing which are locally
jammed. B) Zooming in on the highlighted region in (A), we see that
at the value of $\delta$ where the number of non-rattlers begins
to plataeu, the packing appears to be hypostatic as indicated by a
value of $Z<4$.}

\end{figure}

Having identified rattlers, we replot $Z(\delta)$ with rattlers excluded,
as well as the number of particles which are not identified as rattlers
(fig. \ref{fig:contactsWithoutRattlers}). For $Z$, the intermediate
region is greatly flattened, more closely resembling the plateau observed
in Euclidean space\cite{Donev2005}. Fascinatingly, at $\delta=10^{-6}$,
where the number of non-rattlers begins to plateau, the metric jammed
packing has $Z=3.962$ and has not reached the expected isostatic
value of $Z=4$. As $\delta$ is increased, we do not see $Z=4$ until
$\delta=10^{-5}$, a full order of magnitude higher. 

This is a striking result for two reasons: First, for a system with
such a high degree of hexatic order, we might expect an average contact
number near $Z=6$. We don't see this because curvature of the surface
induces strain in the crystalline regions of the packing\cite{Seung1988}
and the hard particle constraints prevent local compression; rather
the lattice is slightly oblique and hence each particle has only four
contacts in the crystalline regions of the packing. 

\begin{figure}
\begin{centering}
\includegraphics[width=1\columnwidth]{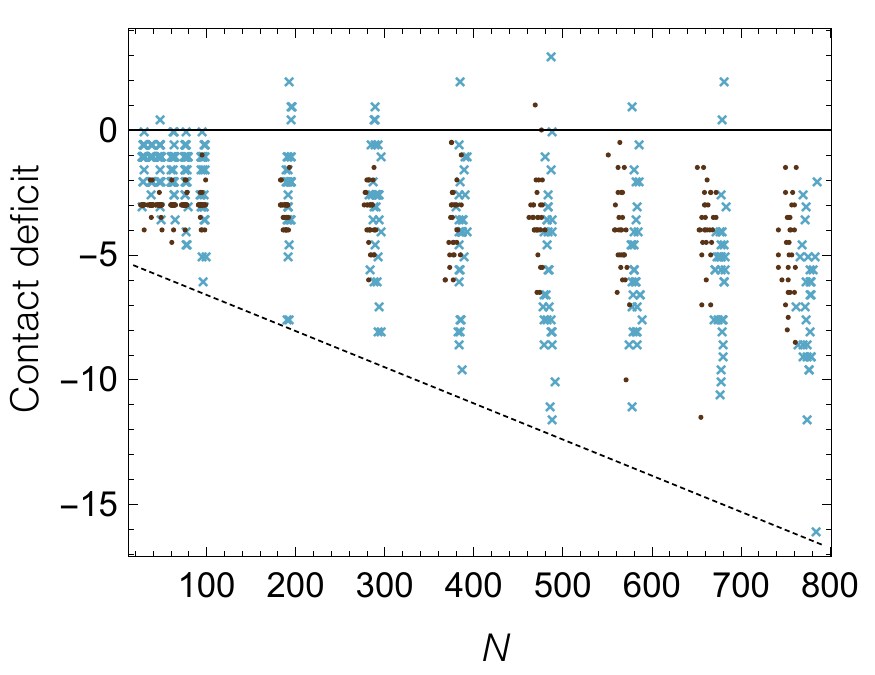}
\par\end{centering}
\caption{\label{fig:Contact-deficit}Contact deficit, the number of contacts
minus the number expected in Euclidean space, for an ensemble of configurations
with different $N$ (not counting rattlers). Blue Xs represent monodispersed
packings and brown points represent bidispersed packings. The dashed
line shows the apparent lower bound. }
\end{figure}

Second, packings with $Z<4$ should be hypostatic according to the
constraint counting argument above. To further explore this, we generated
an ensemble of metric jammed packings at particle number $N$, with
$30$ results at each value. We display the number of missing contacts
in fig. \ref{fig:Contact-deficit}, using $\delta=10^{-6}$ as our
contact cutoff. A few packings are found that have an excess of contacts,
but most display an apparent deficit. Notice that there is an apparent
bound on the number of missing contacts that grows linearly with $N$,
with an approximate slope of $0.014$. We also display data for bidispersed
packings. The particle diameter ratio is $1.4$, a value which results
in a high degree of disorder\cite{OHern2003}. Interestingly, both
monodispersed and bidispersed packings appear to share a similar 
lower bound on the contact number, despite the monodispersed packings
having a high degree of hexagonal ordering compared to the bidisperse
packings.

\begin{figure}
\begin{centering}
\includegraphics[width=3in]{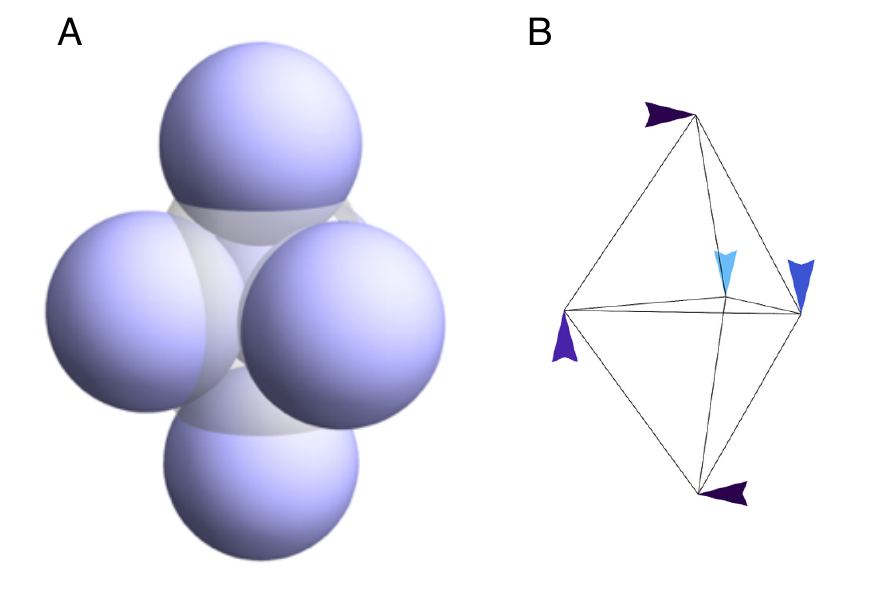}
\par\end{centering}
\caption{\label{fig:ToyExample}\textbf{A} Arrested packing of 5 particles
with the center of masses positioned on the surface of an ellipsoid
of aspect ratio $\sqrt{2}$ at the vertices of a triangular bipyramid.
\textbf{B }The linear programming finds this unjamming motion, representing
rotation about the ellipsoid, but this motion is not feasible within
the nonlinear surface constraint. }
\end{figure}

The reason for the missing contacts is the nonlinear surface constraint.
To show this, we present a toy example shown in fig. \ref{fig:ToyExample}A.
Five spherical particles are arranged at the vertices of a triangular
bipyramid with equilateral faces; these are embedded on a commensurate
ellipsoid of aspect ratio $\sqrt{2}$. The total number of contacts
is $9$; the two particles on the top and bottom touch three other
particles and the three particles around the edge touch four others.
The constraint counting argument in Euclidean space would predict
$10$ contacts were needed for isostaticity. 

Applying the linear program to the example in fig. \ref{fig:ToyExample}A
reveals an apparent unjamming motion shown in fig. \ref{fig:ToyExample}B
that attempts to rotate the configuration about an axis parallel to
the equatorial plane of the ellipsoid. This motion is, however, totally
prevented by the nonlinear surface constraint. The structure is therefore
metric jammed according to the above definition, but it is clear that
the constraint counting for isostaticity must be modified to account
for the curved surface. This is reminiscent of the situation where
non-spherical particles are jammed in Euclidean space\cite{Donev2007}.

\subsection{Soft particles\label{subsec:Soft-particles}}

To generate packings of soft particles, we again employ the dynamic
packing algorithm, replacing the hard particle interactions with a
compact Hertzian interaction,
\[
V_{ij}=\frac{\epsilon}{2/5}\left(1-\frac{r_{ij}}{\sigma_{ij}}\right)^{2/5}\Theta\left(\frac{\sigma_{ij}}{r_{ij}}-1\right)
\]
where $\epsilon$ determines the energy scale, $r_{ij}$ is the distance
between the centroids of particles $i$ and $j$, $\sigma_{ij}$ is
the sum of the radii of the particles, and $\Theta$ is the Heaviside
step function enforcing a finite interaction range. Both monodispersed
packings and bidipsersed packings with a radius ratio of 1.4 are produced.
The surface relaxation proceeds past the point where all particles
are overlapping, creating over-jammed configurations \textemdash{}
the packings become rigid near an aspect ratio of $a=4.0$, and the
surface evolution continues to an aspect ratio of $a=3.0$. All packings
consist of $N=800$ particles. We apply an energy minimization to
the final packing, fixing the surface geometry and using a conjugate
gradient method\cite{numerical_recipes}. From these energy minimized
configurations, we expand the packing quasistatically (i.e. minimizing
the energy after each small expansion step) at fixed aspect ratio.
This allows us to find the jamming point packing fraction $\phi_{c}$
corresponding to the initial energy minimized configuration (as $\phi_{c}$
is a property specific to a given packing, not a universal value\cite{OHern2003}),
while also generating packing fractions in the intermediate range
of packing fractions to study the mechanical behavior of the packings
as a function of $\phi-\phi_{c}$.

First we investigate the vibrational modes of the packings. To calculate
vibrational modes along the surface, we impose a harmonic energy penalty
for particle motions normal to the surface, with an energy scale much
larger than the particle interaction energy scale. We then calculate
the Hessian matrix of the packing energy with respect to the particle
coordinates in 3D and diagonalize it to find eigenfrequencies and
eigenmodes, and ignore modes normal to the surface which are easily
identifiable due to their much higher frequencies.
\begin{figure}
\includegraphics{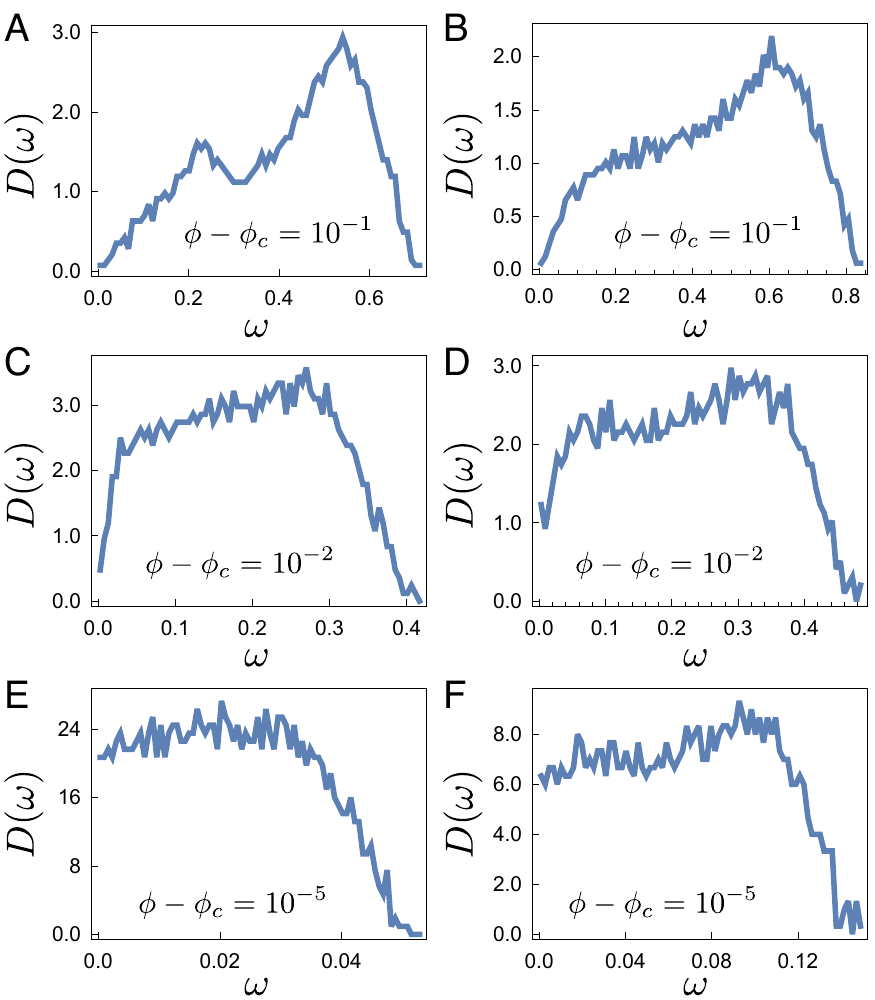}

\caption{\label{fig:vibrationalModes}Density of states of vibrational modes
for a (A,C,E) monodipsersed and (B,D,F) bidispersed packing. In both
cases, as the packing fraction nears $\phi_{c}$, an excess of modes
is observed for at low $\omega$. Notably, the monodispersed packing
does not show Debye law behavior, despite the high degree of hexagonal
order. Frequencies are in units of $\sqrt{\epsilon}/r$, where $r$
is the larger radius for bidisperse packings.}
\end{figure}

Figure \ref{fig:vibrationalModes} shows the density of vibrational
frequencies $D(\omega)$ for both monodispersed and bidispersed packings
at various values of $\phi-\phi_{c}$. We see that as $\phi$ approaches
$\phi_{c}$, the so-called boson peak\cite{Grigera2003}, an excess
of low frequency modes, shifts towards $\omega=0$, and at very low
packings fractions ($\phi-\phi_{c}=10^{-5}$), there is a finite density
of states extending down to $\omega=0$. This is a signature of marginal
stability: below $\phi_{c}$, the particles are not in contact and
all vibrational modes are zero-frequency modes. As a packing approaches
the jamming point from above, it develops an abundance of very low
frequency modes.

\begin{figure}
\includegraphics{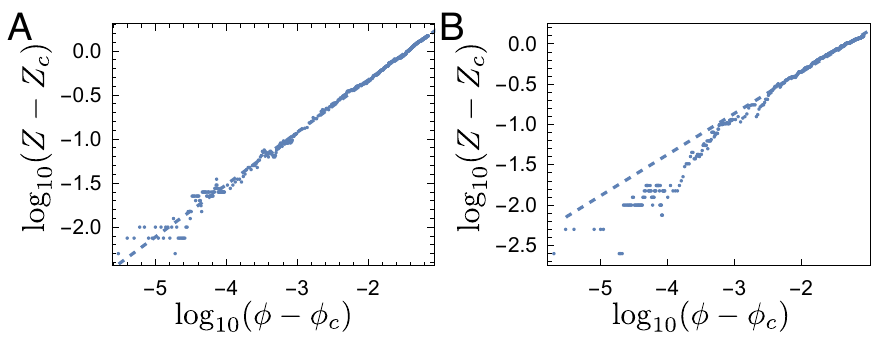}

\caption{\label{fig:softContacts}Scaling of the average contact number versus
packing fraction above the jamming point for (A) a monodispersed packing
and (B) a bidispersed packing. For the monodipsersed case, the data
follows a power law, as shown by the dashed line. For the bidipsersed
case, the data deviates from a power law at lower packing fraction,
but the power law fit for data with $\phi-\phi_{c}>10^{-3}$ describes
the data well.}
\end{figure}

Next we look at the scaling of $Z-Z_{c}$ with respect to $\phi-\phi_{c}$,
where $Z_{c}$ is the contact number at which packings are marginally
stable. For the monodispersed case, we see a power law behavior with
exponent $0.60\pm0.06$, larger than the value of 0.5 usually seen
in disordered packings\cite{OHern2003}. For the bidispersed case,
we find that a power law behavior holds at higher packings fractions:
taking the power law fit for data with $\phi-\phi_{c}>10^{-3}$, we
find a power law exponent of $0.50\pm0.02$, consistent with previous
results for disordered packings. However, the data tends to deviate
form this power law behavior closer to the jamming point. Data for
a single monodispersed and a single bidispersed packing is shown in
fig. \ref{fig:softContacts}. Uncertainties are the standard deviations
in the fit exponents among 30 packings.

Finally, we investigate the scaling of the elastic moduli of packings
near jamming. Disordered jammed systems exhibit a number of nonlinear
elastic behaviors. These can be seen by comparing the instantaneous
response and infinite-time response of their bulk and shear moduli.
The instantaneous moduli are calculated by applying a uniform compression
or shear and then calculating the response of the system pressure.
The infinite-time moduli are calculated by applying the same deformation,
but then minimizing the configuration energy before calculating the
system response. If the system behaves linearly, then the deformation
will scale or shear the configuration's local energy landscape but
will not change its structure and the configuration will still be
at a local energy minimum. In disordered jammed materials, however,
a difference is observed between the instantaneous and infinite-time
response: for the bulk modulus, both the instantaneous response $B_{0}$
and the infinite-time response $B_{\infty}$ show a power law which
scales as $(\phi-\phi_{c})^{\alpha-2}$, where $\alpha$ is the exponent
of the interaction potential (e.g. 5/2 for Hertzian interactions.)
Despite having the same power law exponent, the power laws have different
coefficients such that $B_{\infty}<B_{0}$. The difference is more
extreme for the shear moduli $G_{0}$ and $G_{\infty}$, which have
different power law exponents: $G_{0}\propto(\phi-\phi_{c})^{\alpha-2}$
and $G_{\infty}\propto(\phi-\phi_{c})^{\alpha-1.5}$\cite{OHern2003}.

The bulk and shear moduli can be derived from the pressure tensor.
The pressure tensor of the packing can be calculated by\cite{Allen}
\[
p_{\alpha\beta}=A^{-1}\sum r_{ij\alpha}\frac{r_{ij\beta}}{r_{ij}}\frac{dV}{dr_{ij}}
\]
where $A$ is the surface area, $V$ is the full configuration energy,
and $r_{ij\alpha}$ is the component of $\vec{r}_{ij}$ along the
surface coordinate alpha (we take $\vec{r}_{ij}$ in 3D and take the
projection along the surface tangent vectors $\vec{t}_{\theta}$ and
$\vec{t}_{\phi}$ in the polar and azimuthal directions, respectively,
at both positions $\vec{r}_{i}$ and $\vec{r}_{j}$ and use the average
between the two points.) From this, the bulk modulus can be calculated
from the pressure, $B=\phi\frac{dp}{d\phi}$, where $p=\frac{1}{2}\sum_{\alpha}p_{\alpha\alpha}$
. The shear modulus is given by $G=\frac{d\Sigma}{d\gamma}$ where
$\Sigma=p_{\theta\phi}$ and $\gamma$ is the applied shear. The shear
is applied by twisting the configuration around the ellipsoid symmetry
axis such that $\frac{ds_{\phi}}{ds_{\theta}}$ is constant (where
$s_{\theta}$ and $s_{\phi}$ are arc lengths along the polar and
azimuthal directions), i.e. there is a uniform shear rate across the
surface. After applying a shear we fix the positions of several particles
near the poles of the surface to ensure that the packing will not
relax back completely after the energy minimization, and we exclude
the fixed particles (and the area they cover) from the pressure tensor
calculation. Results for the elastic moduli are given in units of
$\epsilon/r^{2}$ (where $r$ is the larger particle radius in bidispersed
packings.)

\begin{figure}
\includegraphics{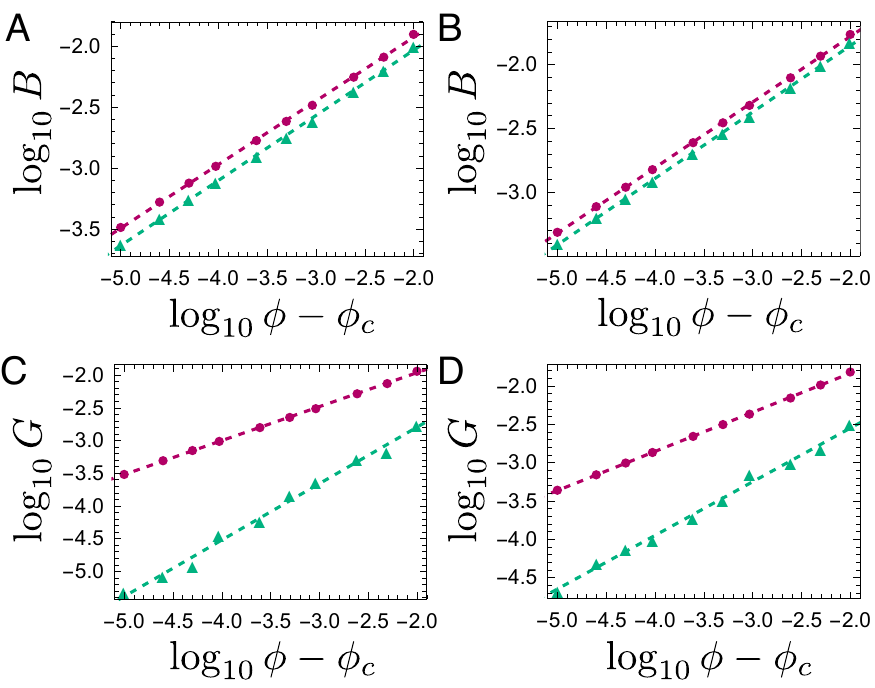}

\caption{\label{fig:moduli}(A, B) Log-log plots of the bulk modulus for a
(A) monodispersed and (B) bidispersed packing. (C, D) Log-log plots
of the shear modulus for a (A) monodispersed and (B) bidispersed packing.
Data for the instantaneous responses $B_{0}$ and $G_{0}$ are shown
as red circles, and the infinite-time responses $B_{\infty}$ and
$G_{\infty}$ are shown as green triangles. Dashed lines are power
law fits. In both A and B, the $B_{\infty}$ is lower than $B_{0}$,
with the same scaling exponent. In C and D, the scaling exponent changes
between $G_{0}$ and $G_{\infty}$. These are all indications of nonlinear
elastic behavior.}
\end{figure}

\begin{table*}
\begin{tabular}{l|c|c|c|c|c}
\hline 
\multicolumn{1}{l}{} & \multicolumn{1}{c}{} & \multicolumn{1}{c}{$B^{0}$} & \multicolumn{1}{c}{$\beta$} & \multicolumn{1}{c}{$G^{0}$} & $\gamma$\tabularnewline
\hline 
\hline 
\multirow{2}{*}{Monodispersed } & Instantaneous & $0.134\pm0.004$ & $0.522\pm0.004$ & $0.123\pm0.005$ & $0.519\pm0.004$\tabularnewline
\cline{2-2} 
 & Infinite-time & $0.111\pm0.007$ & $0.532\pm0.007$ & $0.05\pm0.03$ & $0.72\pm0.07$\tabularnewline
\hline 
\multirow{2}{*}{Bidispersed} & Instantaneous & $0.179\pm0.003$ & $0.514\pm0.002$ & $0.167\pm0.006$ & $0.513\pm0.003$\tabularnewline
\cline{2-2} 
 & Infinite-time & $0.155\pm0.004$ & $0.518\pm0.003$ & $0.08\pm0.03$ & $0.72\pm0.06$\tabularnewline
\hline 
\end{tabular}

\caption{\label{tab:moduliFits}Scaling law fits for the bulk and shear moduli
of both monodispersed and bidispersed packings of 800 particles on
ellipsoids of aspect ratio 3.0. Scaling fits are of the form $B=B^{0}(\phi-\phi_{c})^{\beta}$
and $G=G^{0}(\phi-\phi_{c})^{\gamma}$. For both monodispersed and
bidispersed packings, we see a drop in $B^{0}$ from the instantaneous
to the infinite-time response, while $\beta$ does not change significantly.
Again for both monodispersed and bidispersed packings, we see a change
in $\gamma$ from near 0.5 to 0.72. Both of these behaviors indicate
that monodispersed packings show the same nonlinear behavior as bidispersed
packings. Uncertainties are the standard deviations in the fit parameters
among 30 packings.}
\end{table*}

Table \ref{tab:moduliFits} reports the power law fitting parameters
for the bulk and shear modulus. For the bulk modulus, we see the same
behavior in the relatively well ordered monodispersed packings as
we see in bidipsersed packings: fitting a power law of the form $B=B^{0}(\phi-\phi_{c})^{\beta}$
to the data we see an exponent of about 0.5 (as expected for a Hertzian
potential) for both $B_{0}$ and $B_{\infty}$, and we see that $B_{\infty}<B_{0}$
to a significant degree. For the shear modulus fits of the form $G=G^{0}(\phi-\phi_{c})^{\gamma}$,
we do see a change in the exponent $\gamma$ for both monodipsersed
and bidispersed packings. Curiously, this exponent is about $\gamma=0.72$
for both cases, signifying a smaller change between the instantaneous
and infinite time shear modulus than seen previously in the literature\cite{OHern2003}.
Data and power law fits are shown for a pair of example configurations
in fig. \ref{fig:moduli}.

It is rather surprising that monodipsersed packings on a 2D surface
share so many properties with disordered packings, given that the
monodispersed packings are relatively well ordered. There are two
effects, both stemming from geometric frustration, which may lead
to the packings exhibiting these properties near the jamming point.
First, the surface curvature as well as its topology necessitate topological
defects in the packing\cite{Bowick2000}. These defects correspond
to localized regions of disorder. Second, the curvature causes strain
in the nearly-hexagonal packing\cite{Seung1988}. Thus, instead of
the surface being covered by a perfect hexagonal lattice with each
particle in contact with six neighbors, the lattice is slightly oblique
and most particles have four contacts \textemdash{} allowing for the
average contact number $Z\approx4$ as seen in sec. \ref{subsec:Criteria-for-isotaticity}.

\section{Conclusion\label{sec:Conclusion}}

In section \ref{subsec:Unjamming-arrested-packings} we adapt the
algorithm of \cite{Donev2004b} to search for unjamming motions in
packings of hard particles on curved surfaces. Applying this algorithm
to arrested packings generated by relaxing an ellipsoidal surface
at fixed volume, we find in section \ref{subsec:Are-arrested-packings}
that these arrested packings are generally not jammed. By repeated
unjamming the packings and further relaxing the surface, we artificially
age the packings towards a metric jammed state, that is, the final
packings are stable to both collective particle motions and further
surface evolution.

Upon careful investigation (\ref{subsec:Criteria-for-isotaticity})
we see that the metric jammed packings typically have an average contact
number $Z<4$, where $Z=4$ is the isostatic contact number required
for stability of sphere packings in flat 2D space. This deficit in
the contact number is explained by the surface curvature imposing
nonlinear constraints on the packing, in addition to the constraints
due to interparticle contacts. For packings of increasing particle
number, an increasing number of apparently ``missing'' contacts
is observed. This trend is currently unexplained. Ideally one could
derive an analytical prediction for the number of missing constraints
based on the number of particles, likely from geometric and topological
considerations (perhaps analogous to the result that crystalline packings
on curved surfaces have a net topological defect charge based on their
topology\cite{Hilton1996}).

In section \ref{subsec:Soft-particles}, we turn to packings of soft
particles compressed past the jamming point. By comparing monodispersed
and bidispersed packings, we see that despite the high degree of hexagonal
ordering in monodispersed packings, their mechanical properties near
the jamming point resemble those of highly disordered bidispersed
packings, due to geometric frustration induced by the curved surface.

Another goal for future work would be to further improve the unjamming
linear program for curved surfaces. Possible improvements include
using quadratic constraints, or using a quadratic program instead
of a linear program. This would allow the program to better handle
the nonlinear surface constraints, and thus avoid finding false unjamming
motions. Another important improvement would be to incorporate surface
deformations. Currently, surface evolution and particle unjamming
steps are handled separately. Ideally, they would be combined into
one unjamming program.

\section*{Methods}

\subsection*{Dynamic packing algorithm}

Arrested packings are produced by an algorithm that simulates diffusive
motion of particles on an evolving surface. $N$ particles of fixed
radius $r$ are initially placed with their centers of mass on an
ellipsoidal surface of aspect ratio $a$ by random sequential absorption.
The simulation then proceeds by a sequence of \emph{diffusion steps
}and \emph{surface evolution steps}: diffusion steps evolve the particle
positions according to a Langevin equation,
\[
\vec{x}_{i}^{\prime}(t+\Delta t_{p})=\vec{x}_{i}(t)+\vec{\eta}_{i}\sqrt{2D\Delta t_{p}},
\]
where $\mathbf{\eta}_{i}$ is a random noise term sampled from a gaussian
distribution of unit variance along the tangent plane of the surface
at $\vec{x}_{i}$. Particle centroids are constrained to remain on
the surface by Lagrange multipliers. Surface evolution steps relax
the ellipsoidal surface towards a spherical ground state such that
the area decreases exponentially as a function of simulation time,
\[
A(t)=A_{s}+\left(A_{e}-A_{s}\right)\exp(-\lambda t),
\]
where $A_{s}$ is the area of the final spherical state, $A_{e}$
is the area of the initial ellipsoid and $\lambda$ is the relaxation
constant. Particles are reprojected onto the surface along the normal
direction, and where reprojection causes overlap of particles, a relaxation
step is performed to remove the overlap. To do so, an artificial potential
is applied to the particles,
\[
V_{\mathrm{\mathrm{overlap}}}=\begin{cases}
r^{2}-rx & x<r\\
0 & x\ge r
\end{cases}
\]
and the resulting energy functional is minimized by conjugate gradient.
The simulation employs dynamic time-stepping to ensure accuracy as
the arrest point is approached. 
\begin{acknowledgments}
\emph{The authors wish to thank Aleksandar Donev and Andrea Liu for
helpful discussions. We would also like to thank the Research Corporation
for Science Advancement for financial support through a Cottrell Award. }
\end{acknowledgments}

\bibliographystyle{unsrt}
\bibliography{jammingPaper}

\end{document}